\documentclass[%
 reprint,
 amsmath,amssymb,
 aps,
]{revtex4-2}

\usepackage{graphicx}
\usepackage{dcolumn}
\usepackage{bm}

\usepackage[english]{babel}
\usepackage{float}
\usepackage{blindtext}
\usepackage{soul}
\usepackage[colorlinks=blue, urlcolor=blue, filecolor=blue]{hyperref}
\usepackage{tikz}

\begin{document}

\preprint{APS/123-QED}
\title{Exciton spin structure in lead halide perovskite semiconductors explored via the spin dynamics in magnetic field}

\author{Vladimir L. Zhiliakov$^{1}$, Nataliia~E.~Kopteva$^{2}$, Irina A. Yugova$^{1}$, Dmitri~R.~Yakovlev$^{2,3}$, Ilya~A.~Akimov$^{2,3}$, and Manfred~Bayer$^{2,4}$} 
 
\affiliation{$^{1}$Spin Optics Laboratory, St. Petersburg State University, 198504 St. Petersburg, Russia}
\affiliation{$^{2}$Experimentelle Physik 2, Technische Universit\"at Dortmund, 44227 Dortmund, Germany}
\affiliation{$^{3}$Ioffe Institute, Russian Academy of Sciences, 194021 St. Petersburg, Russia}
\affiliation{$^{4}$Research Center FEMS, Technische Universit\"at Dortmund, 44227 Dortmund, Germany}

\date{\today}

\begin{abstract}
We theoretically investigate the spin structure and spin dynamics of excitons in bulk lead halide perovskite semiconductors with cubic, tetragonal, and orthorhombic crystal symmetry. The exciton spin structure and its modification by an external magnetic field are modeled for different regimes defined by the relative magnitude of the electron–hole exchange interaction (splitting between dark and bright states) and the Zeeman spin splitting. The effects of crystal symmetry and magnetic field orientation with respect to the crystal axes are considered for lead halide perovskite crystals with band gaps in the range $1.4-3.5$~eV, having different ratios of electron and hole $g$-factors.  For cubic symmetry, in a longitudinal magnetic field, our theory predicts quantum beats between the bright exciton states under linearly polarized excitation and detection, while the dark exciton remains optically inactive. In a transverse magnetic field, all exciton spin states become optically active and can be excited by circularly polarized light. Reduction of the crystal symmetry leads to a zero-field offset of the exciton Larmor precession frequencies, modifying the Zeeman splitting energy dependence on magnetic field. This theoretical framework allows for the extraction of the strength of the exchange interaction and the crystal symmetry. Experimentally, we measure the exciton spin coherence via time-resolved photoluminescence at a temperature of 1.6~K in longitudinal and transverse magnetic fields in orthorhombic MAPbI$_3$ crystals. Polarization beats at the frequency of the bright exciton are observed in both configurations. Comparison with theory indicates that the excitons are in the strong exchange interaction regime, and the reduction of symmetry does not lead to a significant splitting of the exciton spin levels.
\end{abstract}

\maketitle

\section{Introduction}

Lead-based perovskites have attracted considerable attention due to their potential applications in optoelectronics, photovoltaics, and spintronics~\cite{Vinattieri2021_book,Vardeny2022_book,Martinez2023_book}. Their relatively low-cost and rapid chemical synthesis enables the formation of a wide variety of structures, including bulk crystals, two-dimensional layers, and nanocrystals. The optical properties of perovskite semiconductors are governed by excitonic effects.

The exciton is formed at the R-point of the Brillouin zone for the cubic lead halide perovskites, and at the $\Gamma$-point for crystals with tetragonal or orthorhombic lattices~\cite{Even2015}. For all symmetries, the perovskites can approximately be considered as model systems for simple two-band semiconductors with the bottom conduction and the top valence band structure being only two-fold spin degenerate. This leads to purely chiral optical selection rules for the excitons in bulk cubic perovskites, as demonstrated in Refs.~\cite{XOO_MAPbI_2025,XOO2024,XOO_Rashba}. It is important to note that the spin properties of bulk lead halide perovskites~\cite{Giovanni2015,odenthal2017,belykh2019,Kopteva_gX_2024} are quite different from those of conventional III-V or II-VI semiconductors with zinc-blende crystal lattice~\cite{excitons:RS_si}. 

The spin structure of an exciton governs its spin dynamics in magnetic field, which are controlled by the interplay of the exciton exchange interaction and the exciton Zeeman splitting~\cite{OO_book, Spin_book_2017}. Magnetic field can efficiently couple the states of the bright exciton triplet~\cite{Tamarat2019_si,Tamarat2023_si}. The basic features of the exciton spin physics for simple two-band semiconductors, like in perovskites, in the presence of a transverse magnetic field were described in Ref.~\cite{excitons:RS_si}, see also Ref.~\cite{Odenthal:2017aa_si}. Here we present the theory of the exciton spin structure in perovskite semiconductors with emphasis on the dynamics of the exciton spin polarization. 

The exciton spin properties in perovskites can be explored using a variety of experimental techniques. Among them, time-resolved pump-probe methods based on Kerr and Faraday rotation are widely used~\cite{belykh2019,kirstein2022am}. More recently, time-resolved differential transmission and reflection spectroscopy have proven their suitability for probing spin dynamics~\cite{giovanni2019,tao2020,bourelle2020}. Yet, in the absence of an external magnetic field, interpreting these measurements is complicated by the spectral overlap of contributions from resident charge carriers and excitons.

For exciton investigation, it is essential to have detailed information on the spin properties of electrons and holes, including their $g$-factors, $g$-factor anisotropies, and spin relaxation mechanisms. In recent years, these properties have been thoroughly determined~\cite{belykh2019,kirstein2022nc,kirstein2022mapi,COO2024}, providing a solid basis for studying excitons in bulk perovskites. In contrast, detailed theoretical and experimental studies of the excitons themselves remain scarce~\cite{XOO_MAPbI_2025,XOO2024}, underlining the need for such an analysis.

Other experimental approaches such as magneto-reflectivity~\cite{Kopteva_gX_2024} and magneto-photoluminescence~\cite{COO2024} under continuous-wave excitation, as well as spin-flip Raman scattering~\cite{kirstein2022nc}, provide valuable insights into exciton spin structure. However, a complete picture of the exciton spin dynamics requires studies of time-resolved photoluminescence~\cite{XOO_MAPbI_2025,XOO2024,XOO_Rashba}. This technique allows one to isolate the exciton contribution both spectrally and temporally, facilitating tracking of the exciton coherent dynamics. Moreover, the application of an external magnetic field helps one to separate overlapping spectral features of electron-hole recombination and exciton emission, enabling a clear identification of the exciton spin dynamics.  Time-resolved photoluminescence has been used to directly monitor the spin precession through exciton quantum beats in the circular and linear polarization degree of the emission in zinc-blende semiconductors~\cite{Heberle1994,Dyakonov1997,Marie2000}, as well as in bulk MAPbI$_3$~\cite{XOO_MAPbI_2025} and FA$_{0.9}$Cs$_{0.1}$PbI$_{2.8}$Br$_{0.2}$~\cite{XOO2024}.

In this paper, we present a comprehensive study of the exciton spin dynamics in bulk lead halide perovskite crystals with cubic, tetragonal, and orthorhombic symmetry. We develop a theoretical model describing the evolution of exciton spin states under the combined action of Zeeman splitting and isotropic electron-hole exchange interaction. Depending on the magnetic field orientation relative to the $\bf{k}$-wavevector of light, the exciton exhibits qualitatively different spin dynamics. In longitudinal magnetic fields, we predict quantum beats of the bright exciton under linearly polarized excitation and detection, with the dark exciton remaining optically inactive. In contrast, in transverse magnetic fields, all exciton spin sublevels become optically active, leading to oscillations in the linear polarization degree of the emitted light. These oscillations occur either at the sum or the difference of the electron and hole Larmor frequencies, governed by the strength of exchange interaction. To validate the theoretical predictions, we perform time-resolved photoluminescence measurements
of an orthorhombic MAPbI$_3$ crystal. The measured exciton dynamics reveal quantum beats consistent with the spin precession of the bright exciton. By comparing experimental data with theory, we identify the regime of strong exchange interaction and extract parameters, such as the bright exciton Land\'e $g$-factor, its anisotropy, the magnitude of the exchange splitting, and its anisotropy induced by the symmetry reduction from cubic.

The paper is organized as follows. In Section~\ref{sec:ESS}, the exciton spin structure in cubic, tetragonal, and orthorhombic lead halide perovskites is introduced. Section~\ref{sec:ESSF} gives the theoretical analysis of the exciton spin structure in longitudinal and transverse magnetic fields. In Section~\ref{sec:ESD},  the exciton spin dynamics in magnetic field are considered. Section~\ref{sec:phases} presents the effect of the symmetry reduction on spin structure and spin dynamics. In Section~\ref{sec:ER}  the experimental results on the exciton spin dynamics measured in orthorhombic MAPbI$_3$ crystal are described. Finally, in Section~\ref{sec:diss}, we compare the theoretical predictions with the experimental observations.

\section{Exciton spin structure}
\label{sec:ESS}

\begin{figure}
    \centering
    \includegraphics[width=1\linewidth]{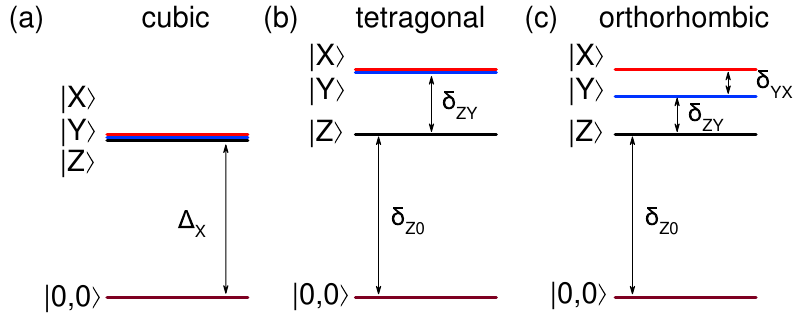}
    \caption{Schematic representation of the exciton spin structure in perovskites with cubic (a), tetragonal (b), and orthorhombic (c) crystal symmetry, shown in the basis of linearly polarized states. $\Delta_{\text{X}}=\Delta_{\text{Y}} < \Delta_{\text{Z}}$ for (b) and $\Delta_{\text{X}}<\Delta_{\text{Y}} < \Delta_{\text{Z}}$ for (c) are assumed.}
    \label{fig:scheme}
\end{figure}

We recall that both the top valence and the bottom conduction bands in the lead halide perovskite semiconductors are two-fold spin degenerate and can be described by spin $1/2$ operators. The Hamiltonian of an electron-hole pair in perovskites with cubic symmetry can be written as:
\begin{equation}
\label{XH}
\hat{H}_\mathrm{exc} = \Delta_\text{X} \hat{\mathbf{s}}_{\rm e} \cdot \hat{\mathbf{s}}_{\rm h}.
\end{equation}
Here, $\Delta_\text{X}$ is the electron-hole exchange splitting, ${\mathbf {\hat s}_\text{e}}$ and ${\mathbf {\hat s}_\text{h}}$ are the electron and hole spin-$1/2$ operators, respectively. The exchange interaction, in agreement with the cubic symmetry, can be recast as
\[
\hat{\mathbf{s}}_{\rm e} \cdot \hat{\mathbf{s}}_{\rm h} = \frac{1}{2} \hat{\mathbf J}^2 - \frac{3}{4},
\]
where $\hat{\mathbf J} = \hat{\mathbf{s}}_\text{e} + \hat{\mathbf{s}}_\text{h}$ is the total spin operator of the electron hole pair and $\hat{\mathbf J}^2 = J(J+1)$. The $J$ takes one of the two values $0$ or $1$.  One can choose the eigenstates of the exchange interaction Hamiltonian in the form $\lvert J,J_z\rangle$ as:
\begin{subequations}
\label{basic:4}
\begin{align}
\label{eq:fi1}
&\phi_1 = \lvert 1,+1\rangle = \lvert \uparrow, \Uparrow \rangle,\\
&\phi_2 = \lvert 1,Z\rangle = -\frac{1}{\sqrt{2}}(\lvert \uparrow, \Downarrow \rangle + \lvert \downarrow, \Uparrow \rangle),\\
&\phi_3 = \lvert 1,-1\rangle = \lvert \downarrow, \Downarrow \rangle,\\
\label{eq:fi4}
&\phi_4 = \lvert 0,0\rangle = \frac{1}{\sqrt{2}}(\lvert \uparrow, \Downarrow \rangle - \lvert \downarrow, \Uparrow \rangle).
\end{align}
\end{subequations}
Here, the up and down arrows indicate the spin orientations $+1/2$ and $-1/2$, respectively. The symbols $\uparrow$ and $\Uparrow$ correspond to electron and hole spins, respectively. In what follows, it is convenient to take the $z$-axis along the $\mathbf k$-vector of light. The states $\phi_1$, $\phi_3$ with angular momentum $J_z= \pm 1$ are optically active in $\sigma^\pm$ circular polarization. The exciton state $\phi_2 = |1,Z\rangle$ has its dipole moment along the $z$-axis (``longitudinal'' exciton). The state $\phi_4 = |0,0\rangle$ is spin forbidden. The exciton eigenstates thus are composed of a spin singlet ($J=0$) and a spin triplet ($J=1$), split by $\Delta_{\rm X}$, as shown in Fig.~\ref{fig:scheme}(a).

Linear combinations of the optically active states for circular polarization are also eigenstates of the exchange Hamiltonian in cubic symmetry:
\begin{equation}
|1,X\rangle=\frac{\lvert \uparrow,\Uparrow\rangle-\lvert\downarrow,\Downarrow\rangle}{\sqrt{2}}, 
\end{equation}
\begin{equation}
|1,Y\rangle=-i\frac{\lvert\uparrow,\Uparrow\rangle + \lvert\downarrow,\Downarrow\rangle}{\sqrt{2}}. 
\end{equation}
These states are optically active in linear polarization.

When lowering the symmetry, we need to consider the Hamiltonian of the exchange interaction in the form:
\begin{equation}
\hat{H}_{\text{exc}}=\sum_{i=\text{X,Y,Z}}\Delta_{i}\hat{\mathbf{s}}_{\text{e},i}\hat{\mathbf{s}}_{\text{h},i}.
\end{equation}
Here $\Delta_{\text{X}}$, $\Delta_{\text{Y}}$ and $\Delta_{\text{Z}}$ are the exchange constants. 
The anisotropy of the exchange interaction leads to splitting of the states, which is shown in Figs.~\ref{fig:scheme}(b,c). For cubic symmetry $\Delta_{\text{X}}=\Delta_{\text{Y}}=\Delta_{\text{Z}}$, for tetragonal symmetry $\Delta_{\text{X}}=\Delta_{\text{Y}}\neq\Delta_{\text{Z}}$, for orthorhombic symmetry $\Delta_{\text{X}}\neq\Delta_{\text{Y}}\neq\Delta_{\text{Z}}$. The eigenstates of the exchange interaction Hamiltonian are $|0,0\rangle$, $|1,X\rangle$, $|1,Y\rangle$ and $|1,Z\rangle$. The energy splittings between these states are determined by the exchange constants:
\begin{eqnarray}
\delta_{\text{YX}}&=&\frac{\Delta_{\text{Y}}-\Delta_{\text{X}}}{2}, \nonumber \\ 
\delta_{\text{ZY}}&=&\frac{\Delta_{\text{Z}}- \Delta_{\text{Y}}}{2}, \nonumber \\  
\delta_{\text{Z0}}&=&\frac{\Delta_{\text{X}}+\Delta_{\text{Y}}}{2}.
\end{eqnarray}

\section{Exciton spin structure in magnetic field for cubic symmetry}
\label{sec:ESSF}

The exciton spin dynamics in an external magnetic field are governed by the interplay between the exchange interaction and the Zeeman splitting $\hat{H}_{\mathrm{Z,X}}$. 
\begin{eqnarray}
\hat{H}&=&\hat{H}_{\mathrm{exc}}+\hat{H}_{\mathrm{Z,X}}.
\end{eqnarray}
The exciton Zeeman splitting is the sum of the electron ($\hat{H}_{\mathrm{Z,e}}$) and hole ($\hat{H}_{\mathrm{Z,h}}$) Zeeman splittings:
\begin{equation}
\label{XH}
\hat{H}_{\mathrm{Z,X}} = \hat{H}_{\mathrm{Z,e}}+\hat{H}_{\mathrm{Z,h}} = (g_\text{e}\hat{\mathbf{s}}_{\rm e}  + g_\text{h}\hat{\mathbf{s}}_{\rm h}) \mu_\text{B}{\bf{B}}.
\end{equation}
Here $\mu_\text{B}$ is the Bohr magneton, $\bf{B}$ is the external magnetic field, $g_\text{e}$ is the electron $g$-factor, and $g_\text{h}$ is the hole $g$-factor. Note that in crystals with cubic symmetry the $g$-factors are isotropic, while their anisotropies need to be accounted for crystals with lower symmetry.

The explicit form of the Hamiltonian in the basis of exciton states $\lvert 0,0\rangle$, $\lvert 1,X\rangle$, $\lvert 1,Y\rangle$, and $\lvert 1,Z\rangle$ for a longitudinal magnetic field (Faraday geometry, $\mathbf{B}_\text{F}\parallel z$ and $\mathbf{k}\parallel z$) reads:
\begin{widetext}
\begin{equation}
   \hat{H}=\frac{1}{2}
    \begin{pmatrix} 0 & 0 & 0 & -g_\text{F,DX}\mu_\text{B}B_\text{F} \\ 
    0 & 2\Delta_{\rm X} & -ig_\text{F,X}\mu_\text{B}B_\text{F} &0 \\ 
    0 & ig_\text{F,X}\mu_\text{B}B_\text{F} & 2\Delta_{\rm X} & 0 \\ 
    -g_\text{F,DX}\mu_\text{B}B_\text{F} & 0 & 0 & 2\Delta_{\rm X} 
    \end{pmatrix},
\end{equation}
\end{widetext}
where $g_\text{F,X} = g_\text{F,e} + g_\text{F,h}$ and $g_\text{F,DX} = g_\text{F,e} - g_\text{F,h}$ are the bright and dark exciton $g$-factors, respectively. The corresponding Larmor precession frequencies are $\omega_\text{F,X}= |g_\text{F,X}|\mu_\text{B}B_\text{F}/\hbar$ and $\omega_\text{F,DX}=  |g_\text{F,DX}|\mu_\text{B}B_\text{F}/\hbar$, where \( \hbar \) is the reduced Planck constant. 

The energies of the exciton states in magnetic field are:
\begin{subequations}
\label{energ:V}
\begin{gather}
\label{eq:EjF1}
E_\text{I} = \frac{1}{2}\left(\Delta_\text{X} - \sqrt{\Delta_\text{X}^2 + (g_\text{F,DX}\mu_\text{B}B_\text{F})^2}\right),\\
\label{eq:EjF2}
E_\text{II} = \frac{1}{2}\left(\Delta_\text{X} + \sqrt{\Delta_\text{X}^2 + (g_\text{F,DX}\mu_\text{B}B_\text{F})^2}\right),\\
\label{eq:EjF3}
E_\text{III} = \Delta_\text{X} - \frac{1}{2}g_\text{F,X}\mu_\text{B}B_\text{F},\\
\label{eq:EjF4}
E_\text{IV} = \Delta_\text{X} + \frac{1}{2}g_\text{F,X}\mu_\text{B}B_\text{F}.
\end{gather}
\end{subequations}

Further, the explicit form of the total Hamiltonian in the basis of exciton states $\lvert 0,0\rangle$, $\lvert 1,X\rangle$, $\lvert 1,Y\rangle$, and $\lvert 1,Z\rangle$ in a transverse magnetic field (Voigt geometry, $\mathbf{B}_\text{V} \parallel x$ and $\mathbf{k}\parallel z$) reads:
\begin{widetext}
\begin{equation}
   \hat{H}= \frac{1}{2}
    \begin{pmatrix} 0 & -g_\text{V,DX}\mu_\text{B}B_\text{V} & 0 & 0 \\
    -g_\text{V,DX}\mu_\text{B}B_\text{V} & 2\Delta_{\rm X} & 0 & 0 \\
    0 & 0 & 2\Delta_{\rm X} & -ig_\text{V,DX}\mu_\text{B}B_\text{V} \\
    0 & 0 & ig_\text{V,DX}\mu_\text{B}B_\text{V} & 2\Delta_{\rm X}
    \end{pmatrix} \,.
\end{equation}
\end{widetext}
As evident from the eigenfunctions of the Hamiltonian presented below (see Eqs.~\eqref{eq:psi34},\eqref{eq:psi12}), all exciton states are optically active in the Voigt geometry. However, we use the notation of bright ($g_\text{V,X} = g_\text{V,e} + g_\text{V,h}$) and dark ($g_\text{V,DX} = g_\text{V,e} - g_\text{V,h}$) exciton $g$-factors by analogy with the Faraday geometry. Here we used the index `V’ to denote the transverse components of the $g$-factors and the corresponding Larmor frequencies ($g_\text{V,i}$ and $\omega_\text{V,i}$) to distinguish from the longitudinal components, which are denoted by the index `F’ ($g_\text{F,i}$ and $\omega_\text{F,i}$), to take into account possible anisotropies of these values in crystals with low symmetry. In particular, as shown in Ref.~\cite{Kopteva_gX_2024}, the bright exciton $g$-factor in bulk lead halide perovskites is about isotropic ($g_\text{V,X} = g_\text{F,X}$), in contrast to $g_\text{DX}$ ($g_\text{V,DX} \neq g_\text{F,DX}$). 

\begin{figure}[b]
    \centering
    \includegraphics[width=0.8\linewidth]{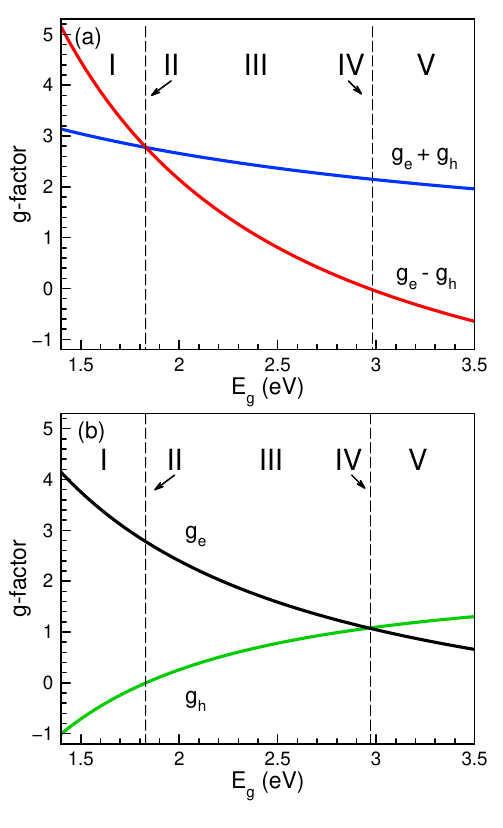}
    \caption{(a) Modeling of the $g$-factor dependence on band gap energy for the bright (blue) and dark (red) excitons in lead halide perovskites, taken from Ref.~\cite{Kopteva_gX_2024}. (b) Hole (green) and electron (black) $g$-factor band gap dependences from Ref.~\cite{kirstein2022nc}. Labels I-V indicate the different cases of $g_\text{e}$, $g_\text{h}$, $g_\text{X}$, and $g_\text{DX}$ values and signs listed in Table~\ref{tab:gF}.}
    \label{fig:EZ}
\end{figure}

\begin{figure*}[t]
    \centering
    \includegraphics[width=1\linewidth]{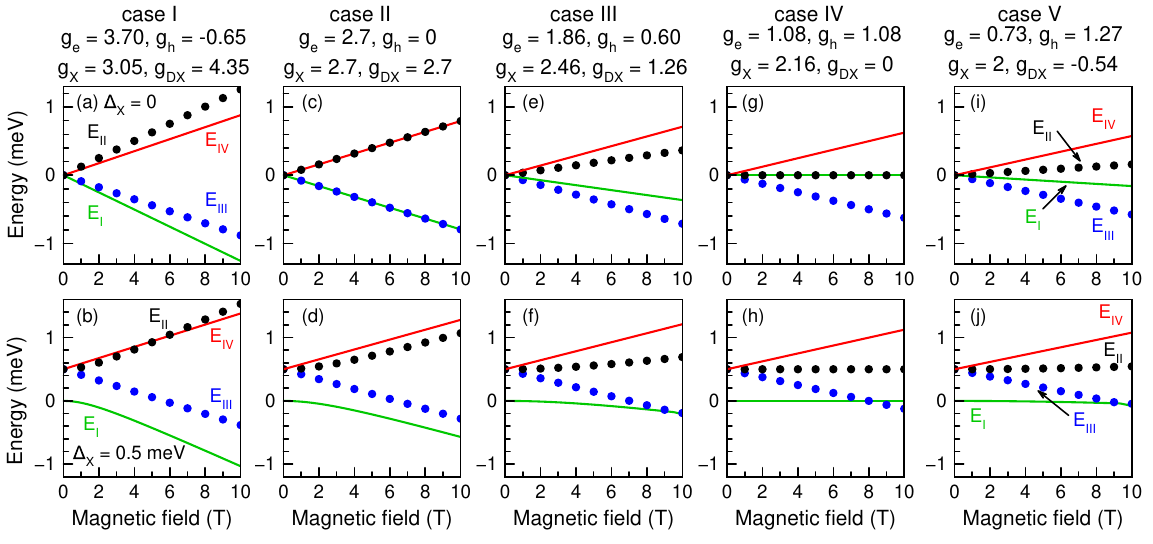}
    \caption{Exciton spin states in magnetic field $B$, calculated for cubic symmetry $\Delta_\text{X} = 0$\,meV (a,c,e,g,i) and $\Delta_\text{X} = 0.5$\,meV (b,d,f,h,j) for different cases of exciton $g$-factors shown in Fig.~\ref{fig:EZ} and listed in Table~\ref{tab:gF}. The $g_\text{e}$, $g_\text{h}$, $g_\text{X}$, and $g_\text{DX}$ values used for the calculations are given on top of the panels. Symbols and lines represent the modeled values for better visualization.}
    \label{fig:AllZ}
\end{figure*}

In the case of an isotropic exchange interaction, the exciton eigenenergies in a transverse magnetic field are identical to those in the longitudinal field configuration, Eqs.~\eqref{eq:EjF1}-\eqref{eq:EjF4}, upon substituting the index F with V. However, all exciton states become optically active for light propagating along the $z$ axis in a transverse magnetic field.

It was recently demonstrated, both experimentally and theoretically, that the electron and hole $g$-factors in lead halide perovskites follow universal dependences on the band-gap energy ($E_\text{g}$)~\cite{kirstein2022nc,arellano2022}. The corresponding calculations are shown in Fig.~\ref{fig:EZ}(b). The contributions to the individual $g$-factors from the $\mathbf{k}\cdot\mathbf{p}$ mixing of these bands proportional to $1/E_\text{g}$ about cancel each other in the bright exciton $g$-factor~\cite{Kopteva_gX_2024}, given by the sum of the electron and hole $g$-factors, as shown by the blue line in Fig.~\ref{fig:EZ}(a). In contrast, the dark exciton $g$-factor calculated as $g_\text{e}-g_\text{h}$ has a strong dependence on $E_\text{g}$, as presented by the red line in Fig.~\ref{fig:EZ}(a). The analysis of the universal trends in the $g$-factors reveals five distinct cases for their relative values and signs, summarized in Table~\ref{tab:gF}. Therefore, we model the exciton spin structure for each of these cases.

Figure~\ref{fig:AllZ} summarizes the dependence of the exciton energy levels on the magnetic field for the cases~I–V and two values of the exchange interaction, \(\Delta_{\rm X}=0\) and 0.5~meV. Note that for the cubic symmetry the exciton energies are independent on the magnetic field geometry. For \(\Delta_{\rm X} = 0\), shown in Figs.~\ref{fig:AllZ}(a,c,e,g,i), all four exciton states exhibit a linear Zeeman splitting, with slopes determined by the corresponding bright exciton (\(g_{\rm X}\)) and dark exciton (\(g_{\rm DX}\)) \(g\)-factors. Variations in \(g_\text{e}\) and \(g_\text{h}\) between the different cases lead to noticeable changes in the slopes of the bright and dark exciton branches. Introducing a finite exchange interaction, \(\Delta_{\rm X} =0.5\)~meV as shown in Figs.~\ref{fig:AllZ}(b,d,f,h,j), keeps the approximate linear field dependence of the state energies~$E_\text{III}$ and~$E_\text{IV}$ over the entire \(B\) range, while the state energies~$E_\text{I}$ and~$E_\text{II}$ show an energy offset at \(B\to 0\) equal to \(\Delta_{\rm X}\), from which a quadratic field dependence emergies at low fields which transforms into a linear dependence only at higher fields. For small values of \(g_{\rm DX}\) (cases IV-V), the spin structure of the dark exciton is governed by the exchange interaction rather than the Zeeman splitting.

\begin{table}
\centering
\caption{Considered ases of $g_\text{e}$, $g_\text{h}$, $g_\text{X}$, and $g_\text{DX}$ values and signs in lead-halide bulk perovskites.}
\begin{tabular}{|c|c|c|c|c|}
\hline
case I & $g_\text{e} > 0$ & $g_\text{h} < 0$ & $g_\text{X} < g_\text{DX}$ & $g_\text{X} > 0$, $g_\text{DX} > 0$ \\ \hline
case II & $g_\text{e} > 0$ & $g_\text{h} = 0$ & $g_\text{X} = g_\text{DX}$ & $g_\text{X} > 0$, $g_\text{DX} > 0$ \\ \hline
case III & $g_\text{e} > 0$ & $g_\text{h} > 0$ & $g_\text{X} > g_\text{DX}$ & $g_\text{X} > 0$, $g_\text{DX} > 0$ \\ \hline
case IV & $g_\text{e} > 0$ & $g_\text{h} > 0$ & $g_\text{X} > g_\text{DX}$ & $g_\text{X} > 0$, $g_\text{DX} = 0$ \\ \hline
case V & $g_\text{e} > 0$ & $g_\text{h} > 0$ & $g_\text{X} > g_\text{DX}$ & $g_\text{X} > 0$, $g_\text{DX} < 0$ \\ \hline
\end{tabular}
\label{tab:gF}
\end{table}

\section{Exciton spin dynamics in magnetic field for cubic symmetry}
\label{sec:ESD}
\subsection{Exciton spin dynamics in longitudinal magnetic field}
A magnetic field modifies the exciton wavefunctions by mixing the eigenstates of the exchange Hamiltonian. In a longitudinal magnetic field (${\bf B}_\text{F} \parallel z$), the bright exciton states are:
\begin{equation}
\psi_\text{III,IV}=\frac{1}{\sqrt{2}}(|1,Y\rangle \pm i|1,X\rangle). 
 \label{eq:p34}
\end{equation}
The corresponding energies \( E_{\text{III,IV}} \) are given by Eqs.~\eqref{eq:EjF3}-\eqref{eq:EjF4}.  

\begin{figure}[b]
    \centering
    \includegraphics[width=1\linewidth]{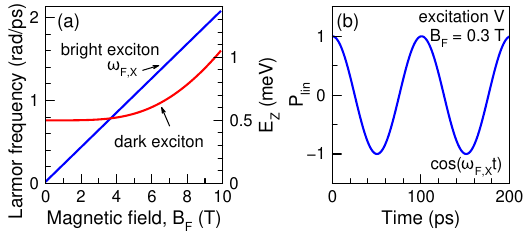}
    \caption{(a) Larmor precession frequency (left axis) and Zeeman splitting energy (right axis) of the bright (blue) and dark (red) exciton in longitudinal magnetic field, $B_\text{F}$. (b) Dynamics of P$_\text{lin}$ at $B_\text{F} = 0.3$~T. $g_\text{F,X} = +2.3$, $g_\text{F,DX} = +2.9$, $\Delta_\text{X} = 0.5$~meV.}
    \label{fig:QBF}
\end{figure}

In this section, we describe the exciton spin dynamics after a short laser pulse. For simplicity, we neglect the finite exciton lifetime and exciton spin relaxation, by considering them infinitely long. Accordingly, the exciton wavefunction \( \Psi(t) \) can be expressed as a coherent superposition of the exciton eigenstates:
\begin{equation}
\Psi(t) = \sum_{i =\text{I}}^{\text{IV}}C_{i} |i\rangle \exp(-i\omega_{i} t), \quad \omega_{i} = E_{i}/\hbar.
\end{equation} 
The coefficients $C_{i}$ are determined by the circular or linear polarization of the laser pulse. Right-circularly polarized light, \( \sigma^{+} =(\mathbf{e}_x + i\mathbf{e}_y) /\sqrt{2}\), 
selectively excites the \( \psi_{\text{III}} \) state, 
whereas left-circularly polarized light, \( \sigma^{-} = (\mathbf{e}_x - i\mathbf{e}_y)/\sqrt{2} \), 
couples exclusively to the \( \psi_{\text{IV}} \) state. Under circularly polarized excitation, circularly polarized emission (as recorded in an optical orientation experiment) of the bright exciton is expected for any $B_\text{F}$.

\begin{figure*}[t]
    \centering
    \includegraphics[width=1\linewidth]{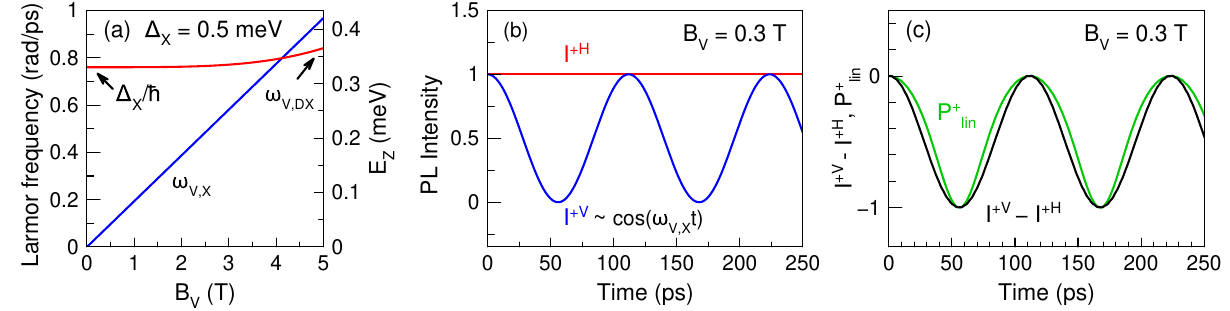}
    \caption{(a) Larmor precession frequencies (left axis) and Zeeman splittings (right axis) of the exciton states in a transverse magnetic field for $\Delta_{\text{X}} = 0.5$~meV. 
(b) Dynamics of $I^\text{+H}$ (red) and $I^\text{+V}$ (blue) after the $\sigma^+$ excitation at $B_{\text{V}} = 0.3$~T.  
(c) Dynamics of linear polarization degree, $P^{+}_{\text{lin}}$ (green), and  $I^\text{+V} - I^\text{+H}$ (black).  $g_\text{V,X} = +2.3$, $g_\text{V,DX} = +3.4$, and $\Delta_\text{X} = 0.5$~meV. }
    \label{fig:QBV_LE}
\end{figure*}

Linearly polarized light excites both the $\psi_\text{III}$ and $\psi_\text{IV}$ states creating their superposition. The dynamics of linear polarization degree of the photoluminescence are oscillating:
\begin{equation}
P^\text{V}_{\mathrm{lin}}(t)=\frac{I^\text{VV}(t)-I^\text{VH}(t)}{I^\text{VV}(t)+I^\text{VH}(t)}\propto \cos(\omega_\text{F,X}t) \,.
\label{eq:BFPlin}
\end{equation}
Here, and throughout, the first superscript denotes the excitation polarization and the second one denotes the detected polarization of the PL. “H” and “V” denote horizontal and vertical linear polarizations. Figure~\ref{fig:QBF}(b) shows that the linear polarization degree \( P_{\mathrm{lin}} \) oscillates at the Larmor precession frequency of the bright exciton.  The frequency \( \omega_\text{F,X} \) depends linearly on the longitudinal magnetic field and does not exhibit a zero field offset.
\begin{equation}
\omega_{\text{F,X}}=|g_{\text{F,X}}|\mu_{\text{B}}B_\text{F}/\hbar\equiv |E_\text{Z}|/\hbar,
\end{equation}
where $E_\text{Z}$ is Zeeman splitting energy (blue line in Fig.~\ref{fig:QBF}(a)).  

In contrast, the Larmor frequency of the dark exciton has a zero-field offset of $\omega_\text{F,DX}(B_\text{F} = 0) = \Delta_\text{X}/\hbar$ determined by the exciton exchange interaction (red line in Fig.~\ref{fig:QBF}(a)). However, in a longitudinal magnetic field, the dark exciton states remain optically inactive. The wavefunctions of the dark exciton states are:
\begin{eqnarray}
\psi_\text{I}=-\sqrt{\frac{Q+\Delta_\text{X}}{2Q}}|0,0\rangle+\sqrt{\frac{Q-\Delta_\text{X}}{2Q}}|1,Z\rangle, \nonumber \\  
\psi_\text{II}=\sqrt{\frac{Q-\Delta_\text{X}}{2Q}}|0,0\rangle+\sqrt{\frac{Q+\Delta_\text{X}}{2Q}}|1,Z\rangle,
\label{eq:p12}
\end{eqnarray}
with the spin splitting of the dark exciton states given by $Q = \sqrt{\Delta^2_\text{X} + (g_\text{F,DX}\mu_\text{B}B_\text{F})^2}$  for the energies $E_\text{I,II}$ given by Eqs.~\eqref{eq:EjF1} and \eqref{eq:EjF2}. 

To summarize, in a longitudinal magnetic field, the exciton spin dynamics addressed by optical methods are governed by the bright exciton states, while the dark exciton states $\lvert 0,0\rangle$ and $\lvert 1,Z\rangle$ do not contribute. 

\subsection{Exciton spin dynamics in transverse magnetic field}

A transverse magnetic field (${\bf B}_\text{V} \parallel x$) mixes the exciton states ($\lvert 1,Y\rangle$, and $\lvert 1,Z\rangle$), modifying the eigenfunctions:
\begin{equation}
\label{eq:psi34}
\psi_\text{III,IV}=\frac{1}{\sqrt{2}}(|1,Z\rangle \pm i|1,Y\rangle). 
\end{equation}
The energies of these states are:
\begin{equation}
\label{eq:EjV1}
E_\text{III,IV}=\Delta_\text{X} \mp \frac{1}{2}g_\text{V,X}\mu_\text{B}B_\text{V}.
\end{equation}
$\lvert 0,0\rangle$ and $\lvert 1,X\rangle$ become also mixed:
\begin{eqnarray}
\label{eq:psi12}
\psi_\text{I} = -\sqrt{\frac{Q+\Delta_\text{X}}{2Q}}|0,0\rangle+\sqrt{\frac{Q-\Delta_\text{X}}{2Q}}|1,X\rangle,\nonumber \\
\psi_\text{II} = \sqrt{\frac{Q-\Delta_\text{X}}{2Q}}|0,0\rangle+\sqrt{\frac{Q+\Delta_\text{X}}{2Q}}|1,X\rangle.
\end{eqnarray}
The energies of these states are:
\begin{equation}
\label{eq:EjV4}
E_\text{I,II}=\frac{1}{2}(\Delta_\text{X}\mp\sqrt{\Delta^2_\text{X} + (g_\text{V,DX}\mu_\text{B}B_\text{V})^2})\equiv\frac{1}{2}(\Delta_\text{X}\mp Q),
\end{equation} 
with $Q = \sqrt{\Delta^2_\text{X} + (g_\text{V,DX}\mu_\text{B}B_\text{V})^2}$. 

\begin{figure*}[t]
    \centering
    \includegraphics[width=1\linewidth]{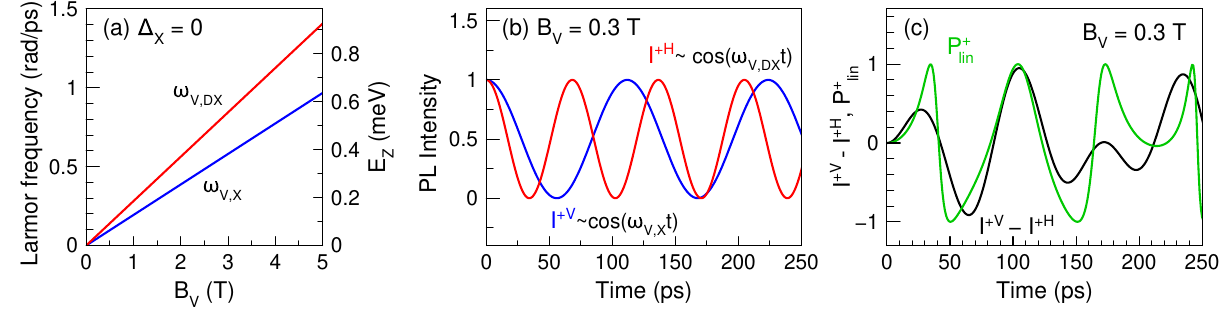}
    \caption{(a) Larmor frequencies (left axis) and Zeeman splittings (right axis) of the exciton states in a transverse magnetic field for $\Delta_{\text{X}} = 0$. 
(b) Dynamics of $I^\text{+H}$ (red) and $I^\text{+V}$ (blue) after $\sigma^+$ excitation at $B_{\text{V}} = 0.3$~T.  
(c) Dynamics of the linear polarization degree, $P^{+}_{\text{lin}}$ (green), and of the intensity difference $I^\text{+V} - I^\text{+H}$ (black). $g_\text{V,X} = +2.3$, $g_\text{V,DX} = +3.4$, and $\Delta_\text{X} = 0$.}
    \label{fig:QBV_SE}
\end{figure*}

\subsection{Strong exchange interaction}
\label{sec:BV_SE}

Let us start considering the exciton spin dynamics in a transverse magnetic field, when the exciton exchange interaction is strong ($\Delta_\text{X} \gg \hbar \omega_\text{V,X}$). The emission intensities polarized perpendicular, $I^\text{+V}$, or parallel, $I^\text{+H}$, to the horizontally-oriented transverse magnetic field after $\sigma^+$ excitation are given by: 
\begin{equation}
I^\text{+V}(t) \propto \frac{1}{2}\left[1 + \cos(\omega_\text{V,X}t\right)],
\end{equation}
\begin{equation}
I^\text{+H}(t) \propto \frac{1}{2}\left(1 + \frac{\Delta^2_{\text{X}}}{Q^2}\right) + \frac{1}{2}\left(1 - \frac{\Delta^2_{\text{X}}}{Q^2}\right) \cos\left(\frac{Q}{\hbar}t\right).
\end{equation}
The symbols “$+$” and “$-$” refer to right and left circular polarizations, respectively. The intensity $I^\text{+V}$ in finite magnetic field exhibits oscillations at the bright exciton precession frequency (blue line in Fig.~\ref{fig:QBV_LE}(b)). The corresponding dependences of the Zeeman splitting and oscillation frequency are shown in Fig.~\ref{fig:QBV_LE}(a). In contrast, the temporal behavior of $I^\text{+H}$ is governed by the strength of the exchange interaction. If $\Delta_\text{X} > E_\text{Z}(B_\text{V} \neq 0)$, the $I^\text{+H}$ component remains non-oscillatory, as illustrated by the red line in Fig.~\ref{fig:QBV_LE}(b).

The resulting dynamics of the linear polarization degree under $\sigma^+$ excitation take the form:
\begin{equation}
P^+_{\text{lin}}(t) = \frac{I^\text{+V}(t)-I^\text{+H}(t)}{I^\text{+V}(t)+I^\text{+H}(t)} \propto \frac{1-\cos{(\omega_\text{V,X}t)}}{3+\cos{(\omega_\text{V,X}t)}}.
\label{eq:BVPlin}
\end{equation}
These dynamics exhibit oscillations at the bright exciton Larmor frequency, as shown by the green line in Fig.~\ref{fig:QBV_LE}(c). For completeness, we also show the dynamics of the intensity difference $I^\text{+V} - I^\text{+H}$ (black line in Fig.~\ref{fig:QBV_LE}(c)), which displays an oscillatory pattern closely resembling that of $P^{+}_{\text{lin}}(t)$.

\subsection{Weak exchange interaction}
\label{sec:BV_WE}

By contrast, let us consider the regime of weak exchange interaction ($\Delta_\text{X} \rightarrow 0$). 
In this limit, the exciton energies are given by: 
\begin{eqnarray}
E_\text{I,II} & \approx & \mp\frac{1}{2} (g_\text{V,DX}\mu_\text{B}B_\text{V}), \\ \nonumber
E_\text{III,IV} & \approx & =\mp\frac{1}{2} (g_\text{V,X}\mu_\text{B}B_\text{V}).
\end{eqnarray}
The corresponding Zeeman splitting energies $E_\text{II} - E_\text{I}$ and $E_\text{IV} - E_\text{III}$ are governed by the dark exciton $\omega_{\text{V,DX}}$ and bright exciton 
$\omega_{\text{V,X}}$ Larmor frequencies, as shown in Fig.~\ref{fig:QBV_SE} by the red and blue lines, respectively.

The exciton wavefunctions are:
\begin{eqnarray}
\psi_\text{I,II}=\frac{1}{\sqrt{2}}(|1,X\rangle \mp |0,0\rangle),\nonumber \\
\psi_\text{III,IV}=\frac{1}{\sqrt{2}}(|1,Z\rangle \pm i|1,Y\rangle). 
\end{eqnarray}
The linearly polarized components of the photoluminescence, $I^\text{+V}$ and $I^\text{+H}$, exhibit oscillations at frequencies corresponding, respectively, to the bright exciton $\omega_\text{V,X}$ and the dark exciton $\omega_\text{V,DX}$ Larmor frequencies:
\begin{equation}
    I^\text{+V}(t) \propto 1+\cos{(\omega_\text{V,X}t)},
\end{equation}
\begin{equation}
    I^\text{+H}(t) \propto 1+\cos{(\omega_\text{V,DX}t)},
\end{equation}
$I^\text{+V}$ and $I^\text{+H}$ are presented in Fig.~\ref{fig:QBV_SE}(b).

The dynamics of the linear polarization degree of the PL excited with $\sigma^+$ polarized light is given by
\begin{equation}
    P^{\text+}_{\text{lin}}(t)\sim\frac{\cos{(\omega_\text{V,X}t)}-\cos{(\omega_\text{V,DX}t)}}{2+\cos{(\omega_\text{V,X}t)}+\cos{(\omega_\text{V,DX}t)}},
\end{equation}
and is illustrated by the green line in Fig.~\ref{fig:QBV_SE}(c). 
Its shows a complex pattern that arises from oscillations of the total intensity, 
$I^\text{+V} + I^\text{+H}$, and the difference $I^\text{+V} - I^\text{+H}$, shown by the black line in Fig.~\ref{fig:QBV_SE}(c).  

\begin{figure*}
    \centering
    \includegraphics[width=0.95\linewidth]{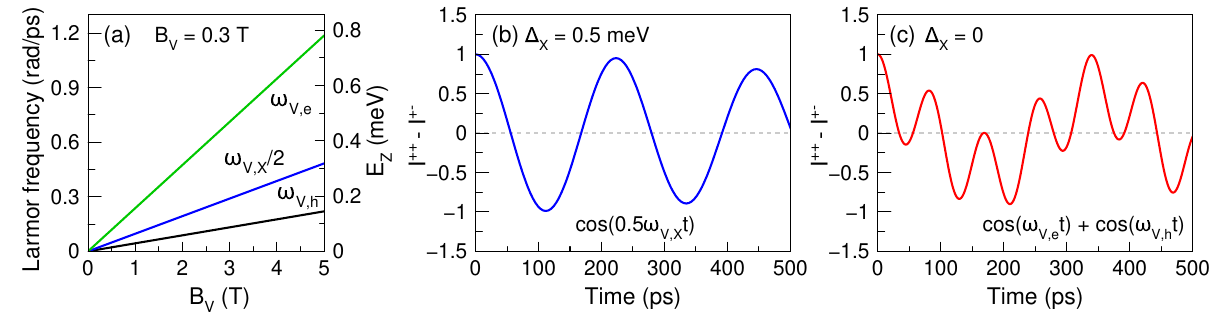}
    \caption{(a) Exciton oscillation frequencies visible in circularly polarized photoluminescence as a function of magnetic field. 
For $\Delta_\text{X} \rightarrow 0$, two oscillation frequencies, $\omega_\text{V,e}$ (green) and $\omega_\text{V,h}$ (black), can appear, whereas for $\Delta_\text{X} = 0.5$\,meV only a single frequency $\omega_\text{V,X}/2$ (blue) appears.
(b,c) Calculated dynamics of $(I^{++} - I^{+-})$ for $\Delta_\text{X} = 0.5$\,meV (b) and $\Delta_\text{X} = 0$\,meV (c) with parameters: $g_{\text{V,e}} = 2.81$ and $g_{\text{V,h}} = -0.68$.} 
    \label{fig:QBV_Circ}
\end{figure*}

\subsection{Optical spin orientation in transverse magnetic field}
\label{sec:BV_Circ}

For completeness of our analysis of the exciton spin dynamics, we consider also the dynamics of optical orientation degree ($P_\text{oo}$):
\begin{equation}
P_{\text{oo}}(t) = \frac{I^{++}(t) - I^{+-}(t)}{I^{++}(t) - I^{+-}(t)},
\label{eq:BVPlin}
\end{equation}
in a transverse magnetic field. This configuration represents the most complex regime, as the dynamics may involve oscillations at up to six distinct Larmor precession frequencies when the exchange interaction is comparable to the Zeeman splitting. However, in the limiting cases of either strong or weak exchange interaction, $P_\text{oo}(t)$ exhibits a simple oscillatory dynamics. If the exchange splitting is large compared to the Zeeman splitting, $\Delta_\text{X} \gg \hbar \omega_\text{V,X}$, the exciton pseudospin precesses with the frequency $\omega_\text{V,X}/2$:
\begin{equation}
P_\text{oo}(t)\sim I^{++} - I^{+-} = 2\cos{\left(0.5\omega_\text{V,X} t\right)},
\end{equation}
as shown in Fig.~\ref{fig:QBV_Circ}(b). Note that the total intensity is not affected by the spin dynamics, and is not considered here.

If the exchange interaction is weak $\Delta_\text{X} \rightarrow 0$, the optical orientation reads:
\begin{equation}
    P_\text{oo}(t) \sim \cos{(\omega_\text{V,e}t)} + \cos{(\omega_\text{V,h}t)}.
\end{equation}
In this regime, the exciton spin dynamics correspond to those of the individual electron and hole spins with their respective Larmor frequencies; see the green and black  lines in Fig.~\ref{fig:QBV_Circ}(a), respectivly. The dynamics of $I^{++} - I^{+-}$ are presented in Fig.~\ref{fig:QBV_Circ}(c). A similar situation occurs for the long-lived time-resolved PL signal originating from electron-hole recombination~\cite{XOO2024,XOO_MAPbI_2025}. 

\section{Exciton Spin Structure in Tetragonal or Orthorhombic Phases}
\label{sec:phases}

\begin{figure}
    \centering
    \includegraphics[width=0.99\linewidth]{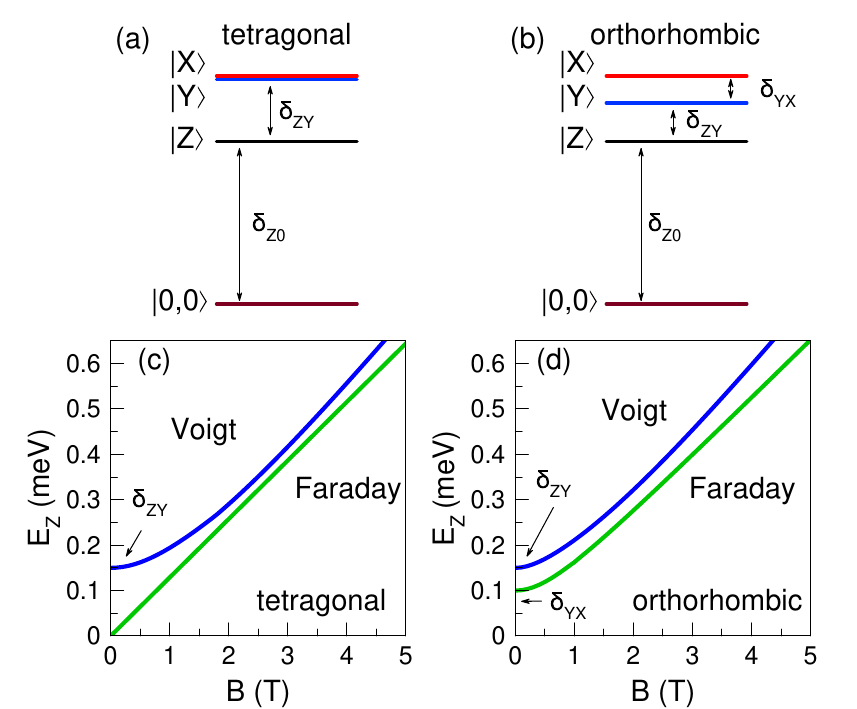}
    \caption{Schematic representation of the exciton spin structure in perovskites with tetragonal (a) and orthorhombic (b) crystal symmetries. Zeeman splitting of the exciton vs. magnetic field in the Faraday (green) and Voigt (blue) geometries for tetragonal (c) and orthorhombic (d) symmetries. Calculation is done with $\delta_\text{YX} = 0.10$~meV, $\delta_\text{ZY} = 0.15$~meV, and $g_\text{V,X} = g_\text{F,X} = +2.3$.}
    \label{fig:Symm}
\end{figure}

Lowering the crystal symmetry to tetragonal or orthorhombic leads to an energy splitting of the exciton triplet states, as illustrated 
schematically in Figs.~\ref{fig:Symm}(a,b). The analysis shows that the mixing of wave functions in the Faraday and Voigt geometries (${\bf B}_\text{F} \parallel z$ and ${\bf B}_\text{V} \parallel x$, $\mathbf{k}\parallel z$) is qualitatively similar to the one in the cubic phase. Here, we consider for simplicity an orientation of the tetragonal or orthorhombic axis (c-axis) along the z-axis ($\mathbf{k}\parallel z \parallel c$). In this case,  $g_\text{F}$ is along the c-axis and $g_\text{V}$ is perpendicular to the c-axis. In the Faraday geometry, the bright and dark excitons are not mixed, although the states in the doublets enter with different coefficients in Eqs.~\eqref{eq:p34},\eqref{eq:p12}. In the Voigt geometry, the states are mixed in pairs $|0,0\rangle$ and $|1,X\rangle$ as well as $|1,Z\rangle$ and $|1,Y\rangle$, similar to Eqs.~\eqref{eq:psi34},\eqref{eq:psi12}, but also with different coefficients compared to these equations.

Further, we focus on calculations of the bright exciton Zeeman splitting in the Faraday and Voigt geometries for the tetragonal and orthorhombic crystal symmetry. In the case of tetragonal symmetry in the Faraday geometry, the resulting Zeeman energy of the bright states varies linearly with magnetic field (the green line in Fig.~\ref{fig:Symm}(c)): 
\begin{equation}
E^{\text{X}}_{\text{Z,F}}= g_\text{F,X}\mu_\text{B}B_\text{F},
\end{equation}
Whereas in the Voigt geometry a finite offset at zero field appears, with magnitude $\delta_{\text{ZY}}$ (the blue line):
\begin{equation}
E^{\text{X}}_{\text{Z,V}}=\sqrt{\delta_{\text{ZY}}^2 + (g_\text{V,X}\mu_\text{B}B_\text{V})^2}.
\end{equation}

In the case of orthorhombic symmetry, all triplet states are split by the 
anisotropic exchange interaction, see Fig.~\ref{fig:Symm}(b). 
In this case, the quantity $\delta_{\text{YX}}$ defines the zero-field offset in the Faraday geometry (the green line in 
Fig.~\ref{fig:Symm}(d)), while $\delta_{\text{ZY}}$ determines 
the corresponding offset in the Voigt geometry (the blue line):
\begin{eqnarray}
E^{\text{X}}_{\text{Z,F}}&=&\sqrt{\delta_{\text{YX}}^2 + (g_\text{F,X}\mu_\text{B}B_\text{F})^2},  \label{eq:orth0}\\
E^{\text{X}}_{\text{Z,V}}&=&\sqrt{\delta_{\text{ZY}}^2 + (g_\text{V,X}\mu_\text{B}B_\text{V})^2}.    \label{eq:orth}
\end{eqnarray}

In strong magnetic fields, the Zeeman splitting in the Faraday and the Voigt geometry follows a linear dependence for both tetragonal and orthorhombic symmetries,  
$E^{\text{X}}_{\text{Z,F(V)}} = g_\text{F(V),X}\mu_\text{B}B_\text{F(V)}$. 

The Zeeman splitting of the other two optically active states in the Voigt geometry is the same for both symmetries:
\begin{equation}
E^{\text{DX}}_{\text{Z,V}}=\sqrt{\delta_{\text{Z0}}^2 + (g_\text{V,DX}\mu_\text{B}B_\text{V})^2}.
\end{equation}

In conclusion, we note that lowering the symmetry results in the appearance of a zero-field offset in the Zeeman splitting. Therefore, a combined analysis of studies in the Faraday and Voigt geometries can be used as a comprehensive tool to determine the crystal symmetry.

\begin{table}
\centering
\caption{ Exciton and charge carrier $g$-factors in bulk MAPbI$_3$ crystals. $g_\text{e}$ and $g_\text{h}$ are taken from Refs.~\cite{kirstein2022mapi,kirstein2022nc}. $|g_\text{X}|$ is measured in this work.}
\begin{tabular}{|c|c|c|c|c|}
\hline
& $g_\text{e}$ & $g_\text{h}$ & $g_\text{e} + g_\text{h}$ & $|g_\text{X}|$  \\ \hline
Faraday geometry& +2.57 & $-0.34$ & +2.23 & 2.2 \\ \hline
Voigt geometry &+2.81 & $-0.68$ & +2.13& 2.5  \\ \hline
\end{tabular}
\label{tab:gMAPbI}
\end{table}


\section{Experimental results}
\label{sec:ER}

As an example of applying the theoretical analysis, we performed time-resolved photoluminescence experiments in a perovskite bulk MAPbI$_3$ crystal with symmetry lower than cubic, at cryogenic temperature.

\subsection{Experimental details}

Single MAPbI$_3$ crystals were grown using the inverse temperature crystallization method~\cite{Chen2019,Alsalloum2020}.  The studied {MA}PbI$_{3}$ single crystal sample has a square shape with dimensions of about 2$\times$2~mm in the (001) crystallographic plane and a thickness of 30~$\mu$m. It has a tetragonal crystal structure at room temperature with an out-of-plane tetragonal $[001]$ axis. At the cryogenic temperatures used in our experiments the crystal structure transforms to orthorhombic. 

In our optical experiments, the geometry with the light wave vector $\textbf{k}\parallel [001]$ was used. The sample was immersed in pumped liquid helium at the temperature of $T=1.6$~K. A superconducting split-coil magnet generates magnetic fields up to 7~T, applied parallel to $\textbf{k}$ (denoted as ${\bf B}_{\rm F}$ in the Faraday geometry) or perpendicular to $\textbf{k}$ (denoted as ${\bf B}_{\rm V}$ in the Voigt geometry).

The photoluminescence is excited by a pulsed Coherent Chameleon Discovery laser. The laser pulses have 100~fs duration at 80~MHz repetition rate and are spectrally tunable from 0.94~eV (1320~nm) to 1.88~eV (660~nm). Time-integrated photoluminescence (PL) and reflectivity spectra were measured with an 0.5\,m spectrometer equipped with a charge-coupled-device (CCD) camera. For measuring recombination and spin dynamics with a time resolution of 6~ps, a streak-camera (nominal time resolution of 2~ps) in combination with an 0.5\,m spectrometer equipped with a 300 grooves/mm diffraction grating was used. To detect exciton spin dynamics, we analyzed the vertically and horizontally linearly polarized components of the photoluminescence. This was done under circularly polarized excitation in the Voigt geometry, and under linearly polarized excitation in the Faraday geometry. Time-integrated PL spectra were obtained by integration of the PL dynamics over time.

\subsection{Optical properties of MAPbI$_{3}$}

\begin{figure}
    \centering
    \includegraphics[width=1\linewidth]{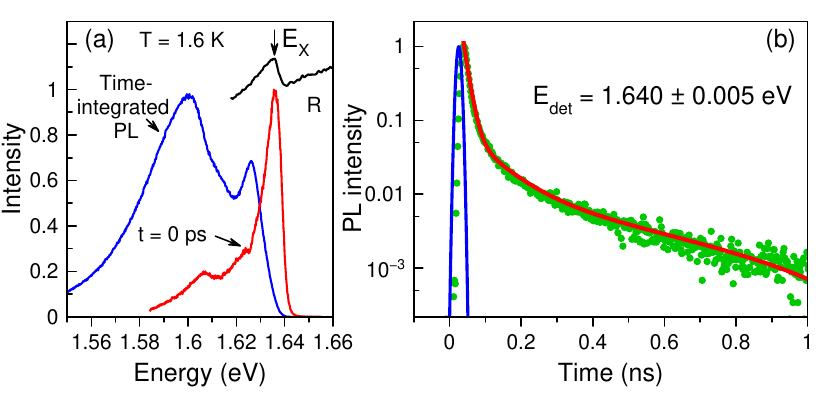}
    \caption{Optical properties of MAPbI$_{3}$ thin crystal measured at $T=1.6$~K.
(a) Reflectivity spectrum (black line) in the vicinity of the exciton resonance $E_\text{X}$ indicated by the arrow. The red line is the PL spectrum right after the laser excitation, taken by time integration over $0-5$~ps. The blue line is the time-integrated PL spectrum. The laser parameters are: $E_\text{exc} = 1.771$~eV with a power density of $P = 10$~mW/cm$^2$. The PL spectra are normalized to their maximal values. (b) Recombination dynamics detected at the exciton resonance and integrated over the spectral range of $1.640\pm 0.005$~eV (circles). The red line is a three-exponential fit with decay times of $\tau_\text{X} = 15$~ps, $\tau_\text{X2} = 85$~ps, and $\tau_\text{R3} = 520$~ps. The laser pulse temporal profile with half-width at half maximum of 6~ps is shown by the blue line.}
    \label{fig:Exp_OP}
\end{figure}

The optical properties of the studied MAPbI$_{3}$ crystal at $T = 1.6$~K are presented in Fig.~\ref{fig:Exp_OP}(a). In the reflectivity spectrum shown by the black line, a pronounced exciton resonance is observed at $E_\text{X} = 1.636$~eV. In time-resolved PL, right after laser pulse excitation, the PL maximum is at the energy of 1.636~eV, corresponding to exciton emission, see the red spectrum in Fig.~\ref{fig:Exp_OP}(a).

The PL dynamics spectrally-integrated in the exciton resonance range $1.640\pm0.005$~eV are shown in Fig.~\ref{fig:Exp_OP}(b). In the temporal range up to 1~ns, the dynamics show a three-exponential decay with two fast decay times of $\tau_\text{X} = 15$~ps and $\tau_{\text{X}2} = 85$~ps, assigned to the exciton recombination. The long component with $\tau_{\text{R}3} = 520$~ps is due to the recombination of spatially separated, localized electrons and holes. This is typical for bulk lead halide perovskite semiconductors, for which the emission lines of electron-hole pairs and excitons are spectrally overlapping~\cite{XOO2024,COO2024,kirstein2022mapi,deQuilettes2019_si,herz2017}. As shown in Ref.~\cite{COO2024}, their contributions can be separated by time-resolved techniques and polarized PL in magneto-optical experiments. More details on the optical and exciton properties of MAPbI$_{3}$ thin crystals can be found in Ref.~\cite{XOO_MAPbI_2025}. 

\subsection{Exciton spin dynamics}

\begin{figure}
    \centering
    \includegraphics[width=0.8\linewidth]{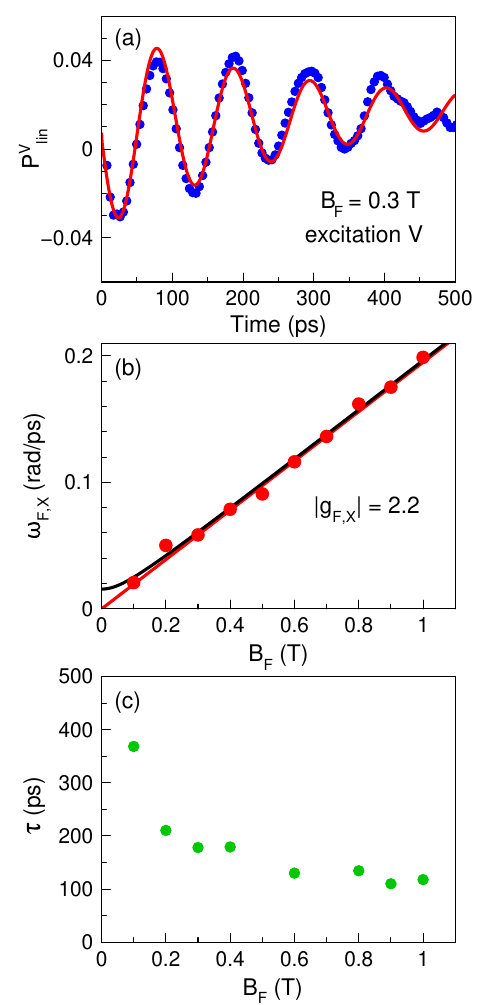}
    \caption{(a) Dynamics of the linear polarization degree of excitons measured in MAPbI$_3$ in the Faraday geometry at $B_\text{F} = 0.3$\,T for vertically (V) polarized excitation (symbols) at $T = 1.6$~K. $E_\text{exc} = 1.771$\,eV with $P = 10$\,mW/cm$^2$. $E_\text{det} = 1.640$~eV. The red line is a fit with Eq.~\eqref{eq:fit}. (b) Experimental dependence of the Larmor precession frequency on $B_\text{F}$ (symbols). The red line is a linear fit with $|g_\text{F,X}| = 2.2$. The black line accounts for the anisotropic exchange interaction in the orthorhombic phase with $\delta_\text{YX}= 10$\,$\mu$eV (Eq.~\eqref{eq:orth0}). (c) Exciton decoherence time $\tau$ as function of $B_\text{F}$.}
    \label{fig:Exp_QBF}
\end{figure}

We focus on the exciton spin dynamics measured by a streak camera to isolate the exciton signal. We measure the vertically $I^\text{VV}$ and horizontally $I^\text{VH}$ polarized components of the photoluminescence in a longitudinal magnetic field $B_\text{F}$, under vertically polarized excitation. The linear polarization degree $P_\text{lin}$ is calculated according to:
\begin{equation}
P^\text{V}_\text{lin}(t) = \frac{I^\text{VV}(t) - I^\text{VH}(t)}{I^\text{VV}(t) + I^\text{VH}(t)}. 
\end{equation}

The dynamics of $P^\text{V}_\text{lin}(t)$ at $B_\text{F} = 0.3$~T are given in Fig.~\ref{fig:Exp_QBF}(a), showing decaying oscillations:
\begin{equation}
\label{eq:fit}
P^\text{V}_\text{lin}(t) = P^\text{V}_\text{lin}(0)\cos(\omega t)\exp{(-t/\tau)}.
\end{equation}
Here $P^\text{V}_\text{lin}(0)$ is the linear polarization degree at zero time delay, $\omega$ is the Larmor precession frequency, and $\tau$ is the decoherence time. The Larmor frequency has a linear dependence on $B_\text{F}$ corresponding to $|g_\text{F,X}| = 2.2$ with no significant zero-field offset, see the symbols in Fig.~\ref{fig:Exp_QBF}(b). This $g$-factor value matches the bright exciton $g$-factor in MAPbI$_3$, obtained from magneto-reflectivity~\cite{Kopteva_gX_2024} and magneto-photoluminescence~\cite{XOO_MAPbI_2025} measurements. These observations support the conclusion that the oscillations in the $P^\text{V}_\text{lin}(t)$ dynamics originate from the bright exciton, while the dark exciton remains optically inactive. This interpretation is confirmed by the theoretical calculations presented in Sec.~\ref{sec:ESD}A.

The decay time of the exciton spin dynamics shown in Fig.~\ref{fig:Exp_QBF}(a), is plotted as a function of magnetic field in Fig.~\ref{fig:Exp_QBF}(c). It significantly exceeds the exciton recombination time, indicating that the decay is governed by the process of  optical decoherence. The decay time decreases with increasing $B_\text{F}$, following a characteristic $1/B_\text{F}$ dependence. 

\begin{figure}
    \centering
    \includegraphics[width=0.8\linewidth]{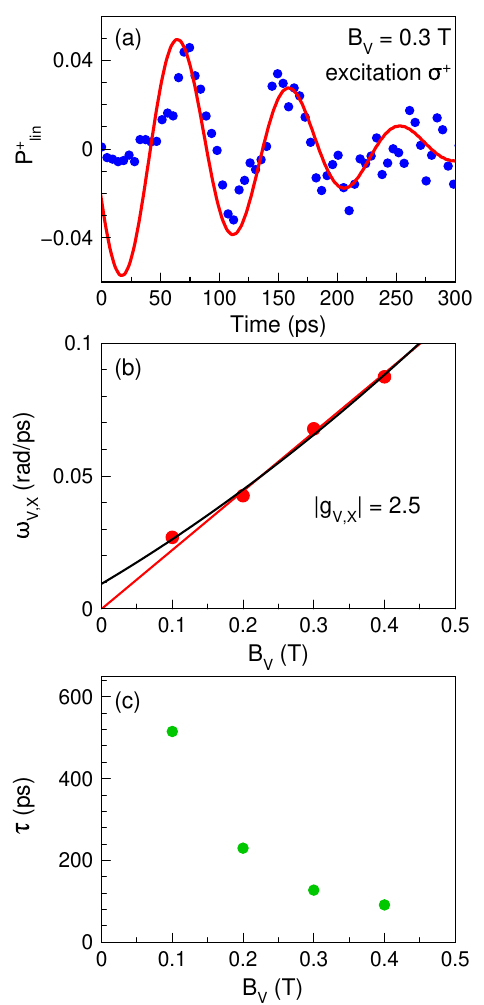}
    \caption{(a) Dynamics of the linear polarization degree of excitons measured in the Voigt geometry at $B_\text{V} = 0.3$\,T for $\sigma^+$ polarized excitation (symbols) at $T = 1.6$~K. $E_\text{exc} = 1.771$\,eV, and $P = 10$\,mW/cm$^2$, $E_\text{det} = 1.640$~eV. The red line is a fit with Eq.~\eqref{eq:fit} with the substitution of index V to index $+$. (b) Experimental dependence of the Larmor precession frequency on $B_\text{V}$ (symbols). A linear fit (the red line) gives $|g_\text{V,X}| = 2.5$. The black line accounts for the anisotropic exchange interaction in the orthorhombic phase with $\delta_\text{ZX}= 7$\,$\mu$eV (Eqs.~\eqref{eq:orth}). (c) Exciton decoherence time ($\tau$) as function of $B_\text{V}$.}
    \label{fig:Exp_QBV}
\end{figure}

In the Voigt geometry, the exciton photoluminescence under circularly polarized excitation becomes linearly polarized. The polarization degree $P^+_\text{lin}$ is calculated according to:
\begin{equation}
P^+_\text{lin}(t) = \frac{I^\text{+V}(t) - I^\text{+H}(t)}{I^\text{+V}(t) + I^\text{+H}(t)}. 
\end{equation}
Here, $I^\text{+V}(t)$ and $I^\text{+H}(t)$ are the vertically and horizontally polarized PL dynamics, perpendicular and parallel to the magnetic field applied in the Voigt geometry (i.e. horizontally). The $P^+_\text{lin}(t)$ dynamics show decaying oscillations, as presented in Fig.~\ref{fig:Exp_QBV}(a). The dynamics  can be fitted with a single-frequency oscillatory function (Eq.~\eqref{eq:fit}). This frequency increases linearly with magnetic field and corresponds to the exciton $g$-factor of $|g_\text{V,X}| = |g_\text{V,e} + g_\text{V,h}|= 2.5$, see Fig.~\ref{fig:Exp_QBV}(b). As discussed in Section~\ref{sec:BV_SE}, this behavior is consistent with the regime of strong exciton exchange interaction. In contrast, if the exchange interaction were weak (see Sec.~\ref{sec:BV_WE}), an additional frequency would be present in the dynamics, reflecting the contribution of the dark exciton states with $|g_\text{V,DX}| = |g_\text{V,e} - g_\text{V,h}|= 3.37$. The values are calculated using the electron and hole $g$-factors of $g_\text{V,e}= 2.83$ and $g_\text{V,h}= -0.54$ as measured in the same MAPbI$_3$ sample~\cite{XOO_MAPbI_2025}. The decay time $\tau$ follows the $1/B_\text{V}$ dependence, as shown in Fig.~\ref{fig:Exp_QBV}(c).


\section{Discussion}
\label{sec:diss}

In bulk perovskite crystals, the 'simple' spin structure of the valence and conduction bands enables the observation of exciton spin beats in the linear polarization degree.

In the Faraday geometry, only spin beats of the bright exciton can be detected, as confirmed experimentally. The beating frequency shows no offset at zero magnetic field, which suggests that the exciton exchange interaction is predominantly isotropic. In contrast, for anisotropic exchange, as expected in the orthorhombic phase, the Larmor precession frequency of the bright exciton would exhibit a finite offset at zero field as illustrated by the black line in Figs.~\ref{fig:Exp_QBF}(b) and \ref{fig:Exp_QBV}(b) for Faraday and Voigt geometry. From these data we estimate for the studied MAPbI$_3$ crystal upper limits of $\delta_\text{ZX} = 7$~$\mu$eV and $\delta_\text{YX} = 10$~$\mu$eV.

In MAPbI$_3$ crystals, a nearly maximal optical orientation degree of 85\% was observed~\cite{XOO_MAPbI_2025}, confirming the unique chiral selection rules. Such a high degree is achievable only if the splitting between the $|X\rangle$ and $|Y\rangle$ states within the triplet is very small. In the opposite case of a large splitting, the emission would be linearly polarized. This provides additional evidence that the anisotropy of the exchange interaction, induced by symmetry lowering, is weak. The Zeeman splitting of the bright exciton in MAPbI$_3$ measured by magneto-reflectivity with high spectral resolution shows no zero field offset confirming that the splitting between the $|X\rangle$ and $|Y\rangle$ states due to the symmetry reduction is very small. In contrast to the excitons, the electrons and the holes are sensitive to symmetry reduction. Their $g$-factors have a strong anisotropy, as was shown by time-resolved Kerr rotation experiments~\cite{kirstein2022mapi}. 

In the Voigt geometry, all exciton states are optically active. For large exchange interaction, the degree of circular polarization exhibits exciton beats at the bright-exciton frequency. The exchange interaction strength controls the amplitude of the dark exciton beats, it reaches maximum for weak exchange interaction. Analysis of these frequencies allows one to estimate the magnitude of the exchange interaction.

The linear polarization beats in the Faraday and Voigt geometries are a clear manifestation of electron-hole spin correlations. An individual charge carrier spin polarization can only produce circularly polarized emission, whereas linear polarization arises from the formation of a correlated electron-hole pair, generated by the same photon. Remarkably, in perovskites the exciton beats can be excited with large detuning from the exciton resonance, in our experiments up to 100~meV. The superposition of exciton spin states can be created by nonresonant excitation. This shows that during energy relaxation, the electron and hole spins photogenerated by the same photon remain correlated and do not lose the coherence of their wave functions. Thus, a significant portion of the excitons relaxes as a whole, rather than as two separate particles. Because of this, resonant and nonresonant excitation affect the exciton in the same way. Note that this is typically not observed in conventional III–V and II–VI semiconductors, where the hole quickly loses its coherence due to spin–orbit interaction and its complex spin structure.
However, further measurements are required to clarify whether exciton formation in perovskite crystals proceeds predominantly via geminate or bimolecular processes.

\section{Conclusions}

We show that the exciton spin dynamics in magnetic field, measured in both circular and linear polarization, provide highly informative insight into the exciton fine structure and its modifications in crystals of various symmetries. Time-resolved photoluminescence, combined with theoretical analysis, opens access to the fine structure energy splittings that cannot be resolved in the spectral domain. We demonstrate this experimentally for a MAPbI$_3$ crystal with orthorhombic symmetry.
The model analysis of the exciton spin structure in magnetic field and the corresponding spin dynamics applies to a broad class of lead halide perovskites with band gaps ranging from 1.4 to 3.5\,eV. The impact of the associated wide variations of the carrier $g$-factors in these materials is discussed. The developed approach can be extended to nanocrystals and 2D materials, where the exchange interaction is strongly modified by quantum confinement and exchange anisotropy.
This study highlights the perovskite semiconductors as a highly attractive platform for exciton and spin physics.

\subsection*{Acknowledgements}
The authors are thankful for fruitful discussions to E.~Yalcin, and B.~F.~Gribakin. N.E.K. acknowledges the support of the Deutsche Forschungsgemeinschaft (project KO 7298/1-1, no. 552699366). I.A.A. acknowledges the support of the Deutsche Forschungsgemeinschaft (AK 40/13-1, no. 506623857).  I.A.Y. and V.L.Z. acknowledge the Russian Science Foundation (Grant no. 19-72-20039) for the financial support of the theoretical analysis performed in this work. 

\section*{Author information}
\textbf{Vladimir L. Zhiliakov} -- orcid.org/0009-0007-0111-9213\\
\textbf{Nataliia E. Kopteva} -- orcid.org/0000-0003-0865-0393\\
\textbf{Irina A. Yugova} -- orcid.org/0000-0003-0020-3679\\
\textbf{Dmitri R. Yakovlev} -- orcid.org/0000-0001-7349-2745\\
\textbf{Ilya~A. Akimov} -- orcid.org/0000-0002-2035-2324\\
\textbf{Manfred Bayer} -- orcid.org/0000-0002-0893-5949\\

\end{document}